\documentclass[a4paper,10pt]{article}

\usepackage[cp1252]{inputenc}
\usepackage[english]{babel}
\usepackage{amsmath,amsfonts,amssymb,amsthm,amstext}
\usepackage{longtable,cite,url,graphicx}
\usepackage{color,relsize}

\newcommand\mysectiona[2]{%
\section
[#1]
{#2}
}

\newcommand\mysectionb[2]{%
\subsection
[#1]
{#2}
}

\newcommand{\notion}[1]{{\bf #1}}
\newcommand{\notioninform}[1]{\text{\bf #1}}

\newcommand{\stdsetnameinform}[1]{%
\mathbb{#1}}

\newtheorem{definition}{Definition}

\newtheorem{theorem}{Theorem}

\newtheorem*{notation*}{Notation}
\newtheorem*{definition*}{Definition}
\newtheorem*{proposition*}{Proposition}
\newtheorem*{lemma*}{Lemma}
\newtheorem*{theorem*}{Theorem}
\newtheorem*{corollary*}{Corollary}

\begin{document}

%
\title{
Notes on space complexity of\\
integration of computable real\\
functions in Ko--Friedman model 
}
%
%
\author{
Sergey\,V.\,Yakhontov
\date{}}

{\def\thefootnote{}
\footnotetext{
\centering
Sergey\,V.\,Yakhontov: Ph.D. in Theoretical Computer Science, Dept. of Computer Science, Faculty of\\
Mathematics and Mechanics, Saint Petersburg State University, Saint Petersburg, Russian Federation, 198504;\\
e-mail: SergeyV.Yakhontov@gmail.com, S.Yakhontov@spbu.ru; phone: +7-911-966-84-30;\\
personal Web page: \url{https://sites.google.com/site/sergeyvyakhontov/; 17-Nov-14}}
%
\maketitle


\abstract{
%
In the present paper it is shown that real function $g(x)=\int_{0}^{x}f(t)dt$ is a linear-space computable
real function on interval $[0,1]$ if $f$ is a linear-space computable $C^2[0,1]$ real function on interval
$[0,1]$, and this result does not depend on any open question in the computational complexity theory.
The time complexity of computable real functions and integration of computable real functions is considered
in the context of Ko--Friedman model which is based on the notion of Cauchy functions computable by
Turing machines.

In addition, a real computable function $f$ is given such that
$\int_{0}^{1}f\in \notioninform{FDSPACE}(n^2)_{C[a,b]}$ but
$\int_{0}^{1}f\notin \notioninform{FP}_{C[a,b]}$ if \notion{FP}$\,\ne\,$\notion{\#P}.
%
}

\vspace{0.2cm}
%

\tableofcontents

%
\mysectiona
{Introduction}
{Introduction}
In the present paper, we consider computable real numbers and functions that are
represented by Cauchy functions computable by Turing machines \cite{K91}.

Main results regarding computable real numbers and functions can be found in \cite{A01,K91,K84,W00};
main results regarding computational complexity of computations on Turing machines can be found in
\cite{DK00}.

As is usual, the set of real functions whose $2$-nd derivative exists and is continuous on interval $[0,1]$
is denoted by $C^{2}[0,1]$, and the set of $C^{k}[0,1]$ real functions for all $k\ge 1$ is denoted by
$C^{\infty}[0,1]$.

It is known \cite{K91} that real function $g(x)=\int_{0}^{x}f(t)dt$ is polynomial-time computable
real function on interval $[0,1]$ iff \notion{FP}$\,=\,$\notion{\#P} wherein $f$ is a polynomial-time
computable real function on interval $[0,1]$. It means integration of polynomial-time computable real
functions is as hard as string functions from complexity class \notion{\#P}.

So this result from \cite{K91} is relativized to question whether \notion{FP}$\,=\,$\notion{\#P}
or not which is one of the open questions in the computational complexity theory.

In the present paper it is shown that real function $g(x)=\int_{0}^{x}f(t)dt$ is a linear-space computable
real function on interval $[0,1]$ if $f$ is a linear-space computable $C^2[0,1]$ real function on
interval $[0,1]$, and this result does not depend on any open question in the computational complexity theory.

%
\mysectionb
{$CF$ computable real numbers and functions}
{$CF$ computable real numbers and functions}
Cauchy functions in the model defined in \cite{K91} are functions binary converging to real numbers.
A function $\phi:\mathbb{N}\rightarrow\mathbf{D}$ (here $\mathbf{D}$ is the set of dyadic rational
numbers) is said to binary converge to real number $x$ if $$|\phi(n)-x|\leq 2^{-n}$$ for all $n\in\mathbb{N}$;
$CF_x$ denotes the set of all functions binary converging to $x$.

Real number $x$ is said to be a $CF$ computable real number if $CF_x$ contains a computable function $\phi$.

Real function $f$ on interval $[a,b]$ is said to be a $CF$ computable function on interval $[a,b]$
if there exists a function-oracle Turing machine $M$ such that for all $x\in[a,b]$
and for all $\phi\in CF_{x}$ function $\psi$ computed by $M$ with oracle $\phi$
is in $CF_{f(x)}$.

The input of functions $\phi$ and $\psi$ is $0^n$ ($0$ repeated $n$ times) when
a number or a function is evaluated to precision $2^{-n}$.  
\begin{definition}
{\normalfont\cite{K91}}
Function $f:[a,b]\rightarrow \stdsetnameinform{R}$ is said to be computable in time $t(n)$ real
function on interval $[a,b]$ if for all computable real numbers $x\in[a,b]$ function $\psi\in CF_{f(x)}$
{\normalfont(}$\psi$ is from the definition of $CF$ computable real function{\normalfont)} is
computable in time $t(n)$.
\end{definition}
\begin{definition}
{\normalfont\cite{K91}}
Function $f:[a,b]\rightarrow \stdsetnameinform{R}$ is said to be computable in space $s(n)$ real
function on interval $[a,b]$ if for all computable real numbers $x\in[a,b]$ function $\psi\in CF_{f(x)}$
{\normalfont(}$\psi$ is from the definition of $CF$ computable real function\normalfont{)} is
computable in space $s(n)$.
\end{definition}
\notion{FP} denotes the class of string functions computable in polynomial time on Turing machines,
\notion{FLINSPACE} denotes the class of string functions computable in linear space on Turing machines, and
\notion{FEXPTIME} denotes the class of string functions computable in exponential time on Turing machines.

According to these notations, polynomial-time computable real functions are said to be
\notion{FP} computable real functions, linear-space computable real functions are said to be
\notion{FLINSPACE} computable real functions, and exponential-time computable real functions are said to be
\notion{FEXPTIME} computable real functions.

The set of \notion{FP} computable real functions on interval $[a,b]$ is denoted by \notion{FP}$_{C[a,b]}$,
the set of \notion{FLINSPACE} computable real functions on interval $[a,b]$ is denoted by
\notion{FLINSPACE}$_{C[a,b]}$, and the set of \notion{FEXPTIME} computable real functions on interval $[a,b]$
is denoted by \notion{FEXPTIME}$_{C[a,b]}$.

The set of the $C^{2}[0,1]$ real functions from class \notion{FP}$_{C[a,b]}$ is denoted by
\notion{FP}$_{C^{2}[a,b]}$, and the set of the $C^{\infty}[0,1]$ real functions from class
\notion{FP}$_{C[a,b]}$ is denoted by \notion{FP}$_{C^{\infty}[a,b]}$.

The same definitions are for other complexity classes.
%
%
\mysectionb
{Integration of \notion{FP} computable real functions}
{Integration of \notion{FP} computable real functions}
The main results from \cite{K91} regarding integration of \notion{FP} computable real functions
are the following.

\begin{theorem}
{\normalfont\cite[5.33]{K91}}
The following are equivalent:
\begin{enumerate}
\item[a)]
{Let $f$ be in \notion{FP}$_{C[0,1]}$. Then, the function $g(x)=\int_{0}^{x}f(t)dt$ is
polynomial-time computable.}
\item[b)]
{Let $f$ be in \notion{FP}$_{C^{\infty}[0,1]}$. Then, the function $g(x)=\int_{0}^{x}f(t)dt$ is
polynomial-time computable.}
\item[c)]
{\notion{FP}$\,=\,$\notion{\#P}.
}
\end{enumerate}
\end{theorem}
It means if \notion{FP}$\,\ne\,$\notion{\#P} then the integral of a polynomial-time computable real function
$f$ is not necessarily polynomial-time computable even if $f$ is known to be infinitely differentiable.
But if $f$ is polynomial-time computable and is analytic on $[0,1]$ then the integral of $f$ must be
computable in polynomial time.

Some additional results regarding the time complexity of integration of computable real functions
can be found in \cite{DK89}. For example, the computation of the volume of a one-dimensional convex
set $K$ is \notion{\#P}-complete if $K$ is represented by a polynomial-time computable function
defining its boundary.
%
%
\mysectiona
{Upper bound of the time complexity of integration}
{Upper bound of the time complexity of integration}
Let $f$ be a linear-space computable $C^2[0,1]$ real function on interval $[0,1]$. Let's consider the composite
trapezoidal rule \cite{BZ62} for function $f$ on interval $[a,x]$ for $a<x\le b$:
\begin{align*}
g(x)&=\int_{a}^{x}f(t)dt=\\
&=\frac{h}{2}\left(f(t_0)+\left(\sum_{i=1}^{k-1}2\cdot f(t_i)\right)+f(t_k)\right)-\\
&\quad-\frac{(b-a)^3}{12\cdot k^2}f''(\xi)
\end{align*}
wherein $k$ is a natural number, $h=\frac{x-a}{k}$, $t_0=a$, $t_i=a+i\cdot h$, $t_k=b$, and $a<\xi<x$.
This equation on interval $[0,x]$ is as follows:
\begin{equation}
\label{Eq:IntegrForm}
\begin{split}
g(x)&=\int_{0}^{x}f(t)dt=\\
&=\frac{h}{2}\left(f(t_0)+\left(\sum_{i=1}^{k-1}2\cdot f(t_i)\right)+f(t_k)\right)-\\
&\quad-\frac{1}{12\cdot k^2}f''(\xi)
\end{split}
\end{equation}
wherein $h=\frac{x}{k}$, $t_0=0$, $t_i=i\cdot h$, $t_k=x$, and $0<\xi<x$.

To compute approximations $g^*(x)$ of function $g$ on interval $[0,1]$, let's compute approximations
$f^*(x_i)$ of function $f$ at points $x_i$, $i\in[0..k]$, to precision $2^{-m}$ wherein $m$ is
a natural number. In that case, the following holds:
\begin{align*}
|\Delta (g; x)|&=|g^*(x)-g(x)|=\\
&=\biggl|\frac{h}{2}\left(f^*(t_0)+\left(\sum_{i=1}^{k-1}2\cdot f^*(t_i)\right)+f^*(t_k)\right)-\\
&\quad-\frac{h}{2}\left(f(t_0)+\left(\sum_{i=1}^{k-1}2\cdot f(t_i)\right)+f(t_k)\right)\biggr|+\\
&\quad+\frac{1}{12\cdot k^2}f''(\xi)\le\\
&\le\frac{h}{2}\biggl(|f^*(t_0)-f(t_0)|+\left(\sum_{i=1}^{k-1}2\cdot |f^*(t_i)-f(t_i)|\right)+\\
&\quad+|f^*(t_n)-f(t_k)|\biggr)+\frac{1}{12\cdot k^2}f''(\xi).
\end{align*}
So, we have:
\begin{align*}
|\Delta (g; x)|&=|g^*(x)-g(x)|\le\\
&\le\frac{h}{2}\left(\frac{1}{2^m}+\frac{2(k-1)}{2^m}+\frac{1}{2^m}\right)+
\frac{1}{12\cdot k^2}C_1\le\\
&\le\frac{1}{k}\frac{k}{2^m}+\frac{1}{12\cdot k^2}C_1=\frac{1}{2^m}+\frac{1}{k^2}C_2;
\end{align*}
here the fact that $f''$ is bounded above if $f\in C^{2}[0,1]$ is taken into account.
If we take $m=2n$ and $k=C_2 2^n$ then
\begin{align*}
|\Delta (g; x)|=\frac{1}{2^{2n}}+\frac{1}{(C_2)^2 2^{2n}}C_2<2^{-n}.
\end{align*}
It means it is sufficient to compute in a loop for $i\in[0,k]$ approximations $g^*(x)$ of function
$g$ on interval $[0,1]$ using formula \eqref{Eq:IntegrForm} wherein $k=C_2 2^n$ and $f^*(x_i)$ are
the approximations of function $f$ at points $x_i$, $i\in[0..k]$, to precision $2^{-m}$
wherein $m=2n$.

So, the following theorems holds.
\begin{theorem}
\label{Th:LinSpIntegr}
If $f$ is a linear-space computable $C^2[0,1]$ real function on interval $[0,1]$ then real function
$g(x)=\int_{0}^{x}f(t)dt$ is a linear-space computable real function on interval $[0,1]$.
\end{theorem}
The same holds for class $\notioninform{FDSPACE}(n^2)$.
\begin{theorem}
\label{Th:n2SpIntegr}
If $f$ is a $\notioninform{FDSPACE}(n^2)$ computable $C^2[0,1]$ real function on interval $[0,1]$ then
real function $g(x)=\int_{0}^{x}f(t)dt$ is a $\notioninform{FDSPACE}(n^2)$ computable real function on
interval $[0,1]$.
\end{theorem}
\begin{theorem}
If $f$ is a polynomial-time and linear-space computable $C^2[0,1]$ real function on interval $[0,1]$
then real function $g(x)=\int_{0}^{x}f(t)dt$ is an exponential-time and linear-space computable real
function on interval $[0,1]$.
\end{theorem}
%
%
\mysectiona
{Function from $\notioninform{FDSPACE}(n^2)_{C[a,b]}$ that not in \notion{FP}$_{C[a,b]}$
if \notion{FP}$\,\ne\,$\notion{\#P}}
{Function from $\notioninform{FDSPACE}(n^2)_{C[a,b]}$ that not in \notion{FP}$_{C[a,b]}$
if \notion{FP}$\,\ne\,$\notion{\#P}}
Let's consider fucntion $f(x)=\sum_{n=1}^{\infty}f_n(x)$ from proof $(d)\Rightarrow(e)$
of theorem 5.32 \cite[p.186]{K91}. Some of the properties of function $f$ are as follows:
\begin{enumerate}
\item[1)]
{$f\in C^{\infty}[0,1]$;
}
\item[2)]
{if $\int_{0}^{1}f\in \notioninform{FP}_{C[0,1]}$ then \notion{FP}$\,=\,$\notion{\#P};
}
\item[3)]
{$f\in \notioninform{FDSPACE}(n^2)_{C[a,b]}$ if one takes set $B$ (\cite[theorem 5.32, p.184]{K91})
such that $B\in \notioninform{DSPACE}(n^2)$;
}
\item[4)]
{$\int_{0}^{1}f\in \notioninform{FDSPACE}(n^2)_{C[a,b]}$ according to theorem \ref{Th:n2SpIntegr}.
}
\end{enumerate}
Therefore, the following theorem holds
\begin{theorem}
$\notioninform{FP}_{C[0,1]}\ne \notioninform{FDSPACE}(n^2)_{C[a,b]}$ if \notion{FP}$\,\ne\,$\notion{\#P}.
\end{theorem}
%
%
\mysectiona
{Conclusion}
{Conclusion}
In the present paper, it is shown that real function $g(x)=\int_{0}^{x}f(t)dt$ is a linear-space
computable real function on interval $[0,1]$ whenever $f$ is a linear-space computable $C^2[0,1]$
real function on interval $[0,1]$. This result differs from the result from \cite{K91} regarding
the time complexity of integration of polynomial-time computable real functions in the sense
that integration of a polynomial-time computable real function \notion{may} be not polynomial-time
computable (if $f$ is not analytic and \notion{FP}$\,\ne\,$\notion{\#P}), but
integration of linear-space computable $C^2$ real functions is \notion{always} linear-space
computable.

Regarding further investigations, it is interesting to derive results regarding the space
complexity of other operators, for example, the space complexity of differentiation of computable
functions. 
%


\end{document}